\documentclass[a4]{amsart}
\usepackage{color}
\usepackage{amsmath}
\usepackage{graphicx}
\usepackage{amsaddr}
\begin{document}
\title{Vortex dynamics and shear layer instability in high intensity cyclotrons}

\author{Antoine J. Cerfon}
\address{Courant Institute of Mathematical Sciences, New York University, New York NY 10012}
\email{cerfon@cims.nyu.edu}
%\date{\vspace{-2ex}}

\begin{abstract}
We show that the space charge dynamics of high intensity beams in the plane perpendicular to the magnetic field in cyclotrons is described by the two-dimensional Euler equations for an incompressible fluid. This analogy with fluid dynamics gives a unified and intuitive framework to explain the beam spiraling and beam break up behavior observed in experiments and in simulations. In particular, we demonstrate that beam break up is the result of a classical instability occurring in fluids subject to a sheared flow. We give scaling laws for the instability and predict the nonlinear evolution of beams subject to it. Our work suggests that cyclotrons may be uniquely suited for the experimental study of shear layers and vortex distributions that are not achievable in Penning-Malmberg traps. 
\end{abstract}

\maketitle
Cyclotrons are efficient and reliable tools for the acceleration of high intensity hadron beams \cite{Seidel}. They are considered a promising option for new applications, including neutrino physics experiments \cite{Daedalus1,Daedalus2} and accelerator driven systems \cite{Harms,Stammbach}. For these high intensity applications, a detailed understanding of the beam dynamics is required to avoid uncontrolled beam loss and activation of the structures. Accordingly, a large part of the theoretical effort has focused on characterizing the influence of space charge on beam quality \cite{Adam,Koscielniak,YangAdel,Bi,Pozdeyev_PRSTAB}. Particle-in-cell (PIC) codes \cite{Hockney} have been predominantly used for that purpose. When combined with supercomputers, they provide quantitative answers that guide the design of machines or help interpret measurements \cite{Bi2}. Even if so, these high performance solvers have large run times, which limits the ability to explore the full range of parameters and configurations, and to identify scaling laws. A complementary approach to study space charge effects is to develop reduced models that retain only the key physical mechanisms to more readily yield scaling laws and physical intuition.

In this letter, we derive a simple fluid model by considering a simplified description of cyclotrons. A self-contained mathematical derivation of the model was given in \cite{CerfonPRSTAB}. Here, we present a more physically intuitive derivation. The assumptions are as follows. First, the confining magnetic field is $\mathbf{B}=B_{0}\mathbf{e}_{z}$, where $B_{0}$ is a constant. Second, we focus on coasting beams, that are not accelerated. The motivation here is to describe a regime that has been extensively studied, precisely to characterize space charge effects \cite{YangAdel,Pozdeyev_PRSTAB}. Our third simplification is to only consider the dynamics in the plane perpendicular to the magnetic field, with physical quantities that only depend on the two coordinates describing that plane. This is reasonable because the physics of a beam immersed in a strong external magnetic field is highly anisotropic. The dynamics along the field couples weakly with the dynamics in the plane perpendicular to it. Lastly, we assume that the motion of the particles is non-relativistic. We do so because the best available data and simulation results correspond to this regime \cite{Adam,YangAdel,Pozdeyev_PRSTAB,PozdeyevThesis}. With these four assumptions, it is natural to work in the frame rotating with the beam at the cyclotron frequency $\omega_{c}=qB_{0}/m$, where $q$ and $m$ are the electric charge and mass of each particle respectively. The equations of motion for the particles in the moving frame are obtained from the equations of motion in the laboratory frame by flipping the direction of the magnetic field $\mathbf{B}=B_{0}\mathbf{e}_{z}\rightarrow \mathbf{B}=-B_{0}\mathbf{e}_{z}$. 
In the moving frame, particles on the equilibrium orbit are static. Particles off this orbit are subject to betatron oscillations \cite{Reiser}, which are periodic circular orbits, whose radius depends on the energy spread, and whose period is $T_{c}=2\pi/\omega_{c}$. Our model relies on the key observation that in cyclotrons the plasma frequency $\omega_{p}=(q^2N_{0}/\epsilon_{0}m)^{1/2}$, where $N_{0}$ is the peak number density of the beam and $\epsilon_{0}$ is the vacuum permittivity, is smaller than the cyclotron frequency $\omega_{c}$. Most machines satisfy $\delta^2\equiv\omega_{p}^2/\omega_{c}^2\ll 1$. When $\delta\ll 1$, the motion of the charged particles consists of the periodic betatron motion plus a small perturbation due to space charge forces. We use this fact to average the equations for the beam dynamics over the betatron time scale, and obtain a description that is relevant on the slow space charge time scale. The last important ingredient in our theory is to focus on the space-charge dominated regime, in which the Debye length is smaller than the characteristic size of the beam $a$. This requires that the radius of the betatron orbits in the moving frame is of order $\delta a$.

In the regime described above, fluid equations derived from a small gyroradius ordering are valid \cite{HazeltineMeiss}. Conservation of mass takes the usual form:
\begin{equation}\label{eq:mass_conservation}
\frac{\partial n}{\partial t}+\nabla\cdot\left(n\mathbf{V}\right)=0
\end{equation}
where $n$ is the number density, and $\mathbf{V}$ is the fluid velocity perpendicular to the magnetic field. To lowest order in $\delta$, $\mathbf{V}$ is given by the sum of the $\mathbf{E}\times\mathbf{B}$ drift and the diamagnetic drift, which is the net cross-field fluid velocity associated with a pressure gradient\cite{HazeltineMeiss,Montgomery}:
\begin{equation}\label{eq:cross_velocity}
n\mathbf{V}=\frac{1}{m\omega_{c}}\left(qn\nabla\phi+\nabla p\right)\times\mathbf{e}_{z}
\end{equation}
where $\phi$ is the electrostatic potential, defined by $\mathbf{E}=-\nabla\phi$, and $p$ is the fluid pressure. Inserting Eq.\eqref{eq:cross_velocity} into Eq.\eqref{eq:mass_conservation}, we obtain the equation for the evolution of the beam density on the space charge time scale
\begin{equation}\label{eq:final1}
\frac{\partial n}{\partial t}+\frac{q}{m\omega_{c}}\nabla\phi\times\mathbf{e}_{z}\cdot\nabla n=0
\end{equation}
The electrostatic potential $\phi$ in Eq.\eqref{eq:final1} is calculated through Poisson's equation: $\nabla^2\phi=-qn/\epsilon_{0}$. The last step is to nondimensionalize the equations, through the rescalings $\overline{n}=n/N_{0}$, $\overline{t}=\omega_{c}t$, $\overline{\nabla}=a\nabla$,$\overline{\phi}=\epsilon_{0}/qa^2N_{0}\phi$. Dropping the bars over the nondimensional quantities, we obtain the desired nondimensional form:
\begin{equation}\label{eq:nondimensional}
  \begin{split}
    \frac{\partial n}{\partial t}+\delta^2\nabla\phi\times\mathbf{e}_{z}\cdot\nabla n=0 \\
    \nabla^2\phi=-n
  \end{split}
\end{equation}
Eq.\eqref{eq:nondimensional} has a simple interpretation: on the space charge time scale, the beam evolves due to advection by the $\mathbf{E}\times\mathbf{B}$ flow field. Despite its simplicity, this model describes a variety of beam phenomena and agrees closely with experimental and numerical results.
The first consequence of Eq.\eqref{eq:nondimensional} is that the strength of the space charge effects appears only through $\delta^{2}$. Since $\delta^{2}$ can be eliminated from Eq.\eqref{eq:nondimensional} by a rescaling of the time variable, the beam density only affects the time scale over which phenomena will be observed, but not their nature. Furthermore, the growth rates of instabilities must depend linearly on $\delta^2$, in agreement with experimental results and simulations \cite{Pozdeyev_PRSTAB}. This scaling required modifications of existing models \cite{Bi,Pozdeyev_PRSTAB}, but is an immediate consequence in our model. The second remarkable point is that Eq.\eqref{eq:nondimensional} is formally identical to the vorticity-stream function form of the two-dimensional Euler equations for an incompressible fluid. In the Euler case, the vorticity plays the role of the density $n$, and the stream function plays the role of the electrostatic potential $\phi$. This isomorphism has been highlighted in a different context \cite{Driscoll,Fine}. A major difference is that in \cite{Driscoll,Fine}, the plasma is surrounded by a cylindrical conductor, whereas in our model Poisson's equation is solved with free space boundary conditions. Some of the phenomena we discuss in this letter occur in elongated beams that would be hard to create in the experiments described in \cite{Driscoll,Fine} and have not been studied in that context. The isomorphism between our model and the Euler equations allows us to study the beam dynamics by relying on the rich fluid dynamics literature and on available numerical solvers. This is what we do in the remainder of this letter. Our numerical results were obtained with a code we adapted from a freely available Euler code based on upwinding \cite{Roullet}, in which we changed the boundary conditions and the Poisson solver \cite{Shidong} in order to impose free space boundary conditions. 

\begin{figure*}
\includegraphics[width=\linewidth]{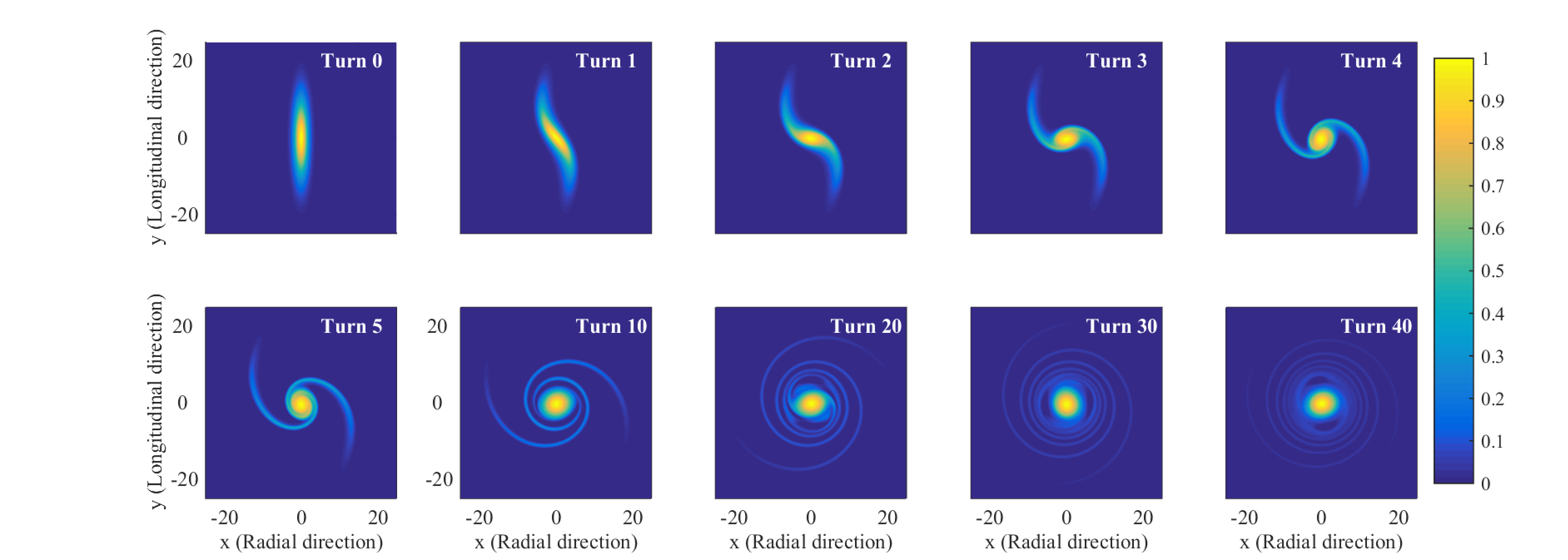}
\caption{\textbf{Spiraling and axisymmetrization of an elongated beam} (color). The dynamics of the beam is given by Eq. \eqref{eq:nondimensional}, and the initial density distribution is $n(x,y)=\exp(-x^2/2\sigma_{x}^2-y^2/2\sigma_{y}^2)$, with $2\sigma_{x}=2.52$, $2\sigma_{y}=13.4$, and $\delta^2=0.8$. The beam propagates in the negative $y$ direction.}
\label{fig:spiral}
\end{figure*}

We start by noting that round beams, such that $n=n(\sqrt{x^2+y^2})=n(r)$, are exact solutions of Eq.\eqref{eq:nondimensional}, regardless of the functional dependence on $r$. Indeed, in that case the $\mathbf{E}\times\mathbf{B}$ drift is everywhere tangential to the density contours. Round beams are also dynamically important, in that they are the final state of a wide class of initial distributions. This explains why round beams are experimentally found to be stable and long lived, and are the final state of simulations \cite{Adam,Koscielniak,YangAdel}. To understand this, let us consider the oft-studied case \cite{Adam,Koscielniak,YangAdel} of the Gaussian density distribution $n(x,y)=\exp(-x^2/2\sigma_{x}^2-y^2/2\sigma_{y}^2)$, where $\sigma_{x}$ and $\sigma_{y}$ are the variances in the $x$ and $y$ directions. In fluid dynamics, such vorticity distributions are known to be subject to ``axisymmetrization" \cite{Melander,Balmforth}: the sheared $\mathbf{E}\times\mathbf{B}$ flows are strongest at the extremities of the beam, leading to the formation of spiral arms which break the elliptic symmetry. The spiral arms get wound up around the beam core, eventually leading to an axisymmetrized beam distribution. This is what we show in Figure \ref{fig:spiral}, obtained for parameters relevant to the PSI Injector-II cyclotron \cite{YangAdel}: $2\sigma_{x}=2.52$, $2\sigma_{y}=13.4$, and $\delta^2=0.8$. The good agreement between our reduced model and the PIC simulations \cite{Adam,Koscielniak,YangAdel} is remarkable considering that $\delta^2=0.8$, and that references \cite{Adam,Koscielniak,YangAdel} included the geometric variations of the magnetic field. The latter can be explained by the rapid motion of the particles that smears out the details of the magnetic field on the slow space charge time scale. Note that a significantly more accurate numerical scheme is used here as compared to the similar computation reported in \cite{CerfonPRSTAB}. The figure shown here thus follows true Euler dynamics more closely. Conservation laws for Euler's equations tell us that according to Eq. \eqref{eq:nondimensional} the spiral arms do never decay, and axisymmetrization remains incomplete for sharp beam distributions \cite{Balmforth}. However, in fluids small amounts of viscosity are sufficient to sustain the axisymmetrization process until one obtains a round beam \cite{Balmforth}. This is what happens in numerical simulations, due to numerical noise and truncation errors. This is also likely the fate of the spiral halo in experiments. Once the width of filaments is much smaller than $a$, our reduced model is no longer justified, and additional terms should be included, such as the nondiagonal components of the pressure tensor, which would provide dissipation at these scales\cite{Driscoll}. 
%From a practical point of view, two observations can be made: 1) because of spiraling, the radial extent of the beam halo is comparable to the original longitudinal length of the beam, which has direct consequences for beam extraction and for the interaction with bunches on neighboring turns; 2) even if $\delta^2$ is relatively large, so that axisymmetrization can be seen as a desirable effect occuring in the first few turns of the cyclotron allowing the spiral halo to be intentionally stripped, the final round beam has a radial distribution that is wider than the original radial distribution.

\begin{figure}
\includegraphics[width=\linewidth]{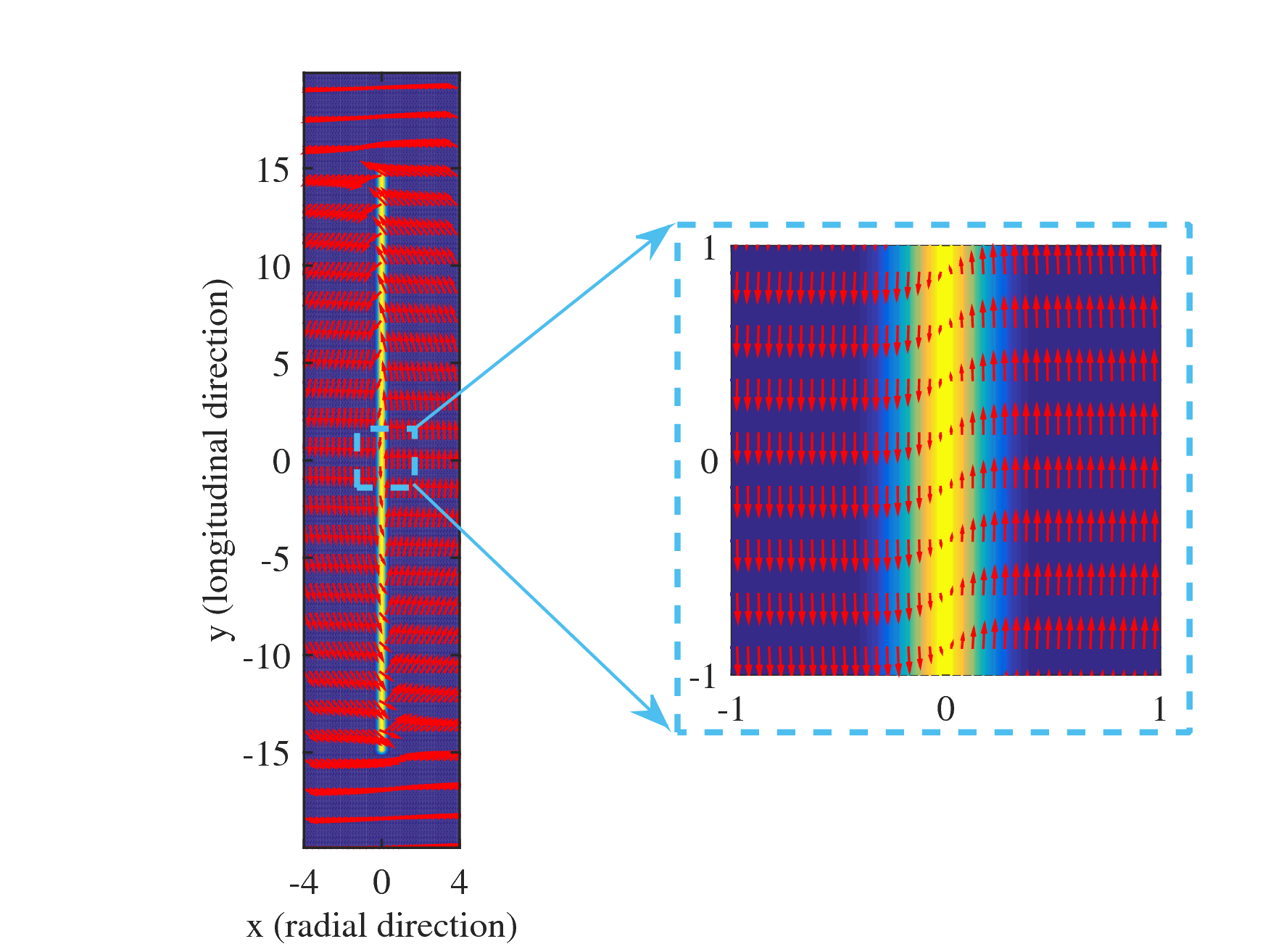}
\caption{$\mathbf{E}\times\mathbf{B}$ \textbf{velocity field for a highly elongated SIR beam} (color). The velocity field is represented by red arrows, and was computed for the density distribution $n(x)=\exp(-\beta x^2)$ if $-15\leq y\leq 15$ and $n(x)=0$ otherwise, with $\beta=-4\ln(10^{-3})$ so that $n=10^{-3}$ at $x=\pm 1/2$.}
\label{fig:shearlayer}
\end{figure}

\begin{figure}
\includegraphics[width=\linewidth]{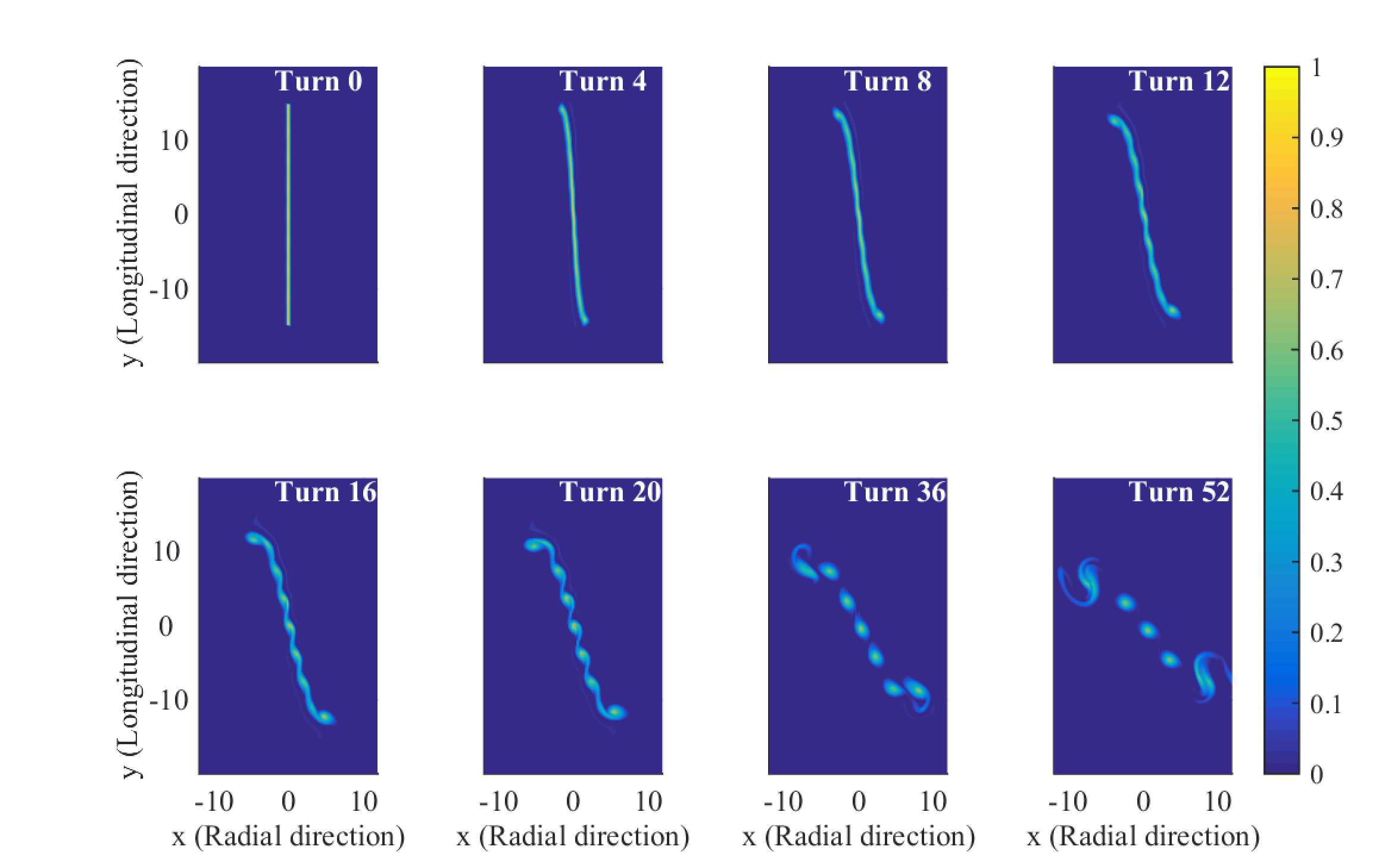}
\caption{\textbf{Break up of the beam shown in Fig. \ref{fig:shearlayer}} (color). We added a small periodic perturbation to the initial beam profile, in order to trigger the instability: for $-15\leq y\leq 15$, the initial beam distribution is given by $n(x)=\exp(-\beta x^2/h_{p}^{2})$, with $\beta=-4\ln(10^{-3})$, and the perturbed layer width is $h_{p}=1+\alpha\cos(ky)$, with $\alpha=.025$ and $k=1.68$. The evolution of the beam is determined by solving Eq. \ref{eq:nondimensional}. Turn 36 and 52 show the development of vortex mergers leading to the formation of a single large round cluster.}
\label{fig:instability}
\end{figure}

In the last part of this letter, we look at the break up instability of elongated beams \cite{Bi,Pozdeyev_PRSTAB,PozdeyevThesis}. It has been identified in PIC simulations and in experiments on the Small Isochronous Ring (SIR) \cite{Pozdeyev_PRSTAB,PozdeyevThesis}, a ring that can operate in the isochronous regime and emulate the beam dynamics of high intensity cyclotrons. We follow two approaches to characterize the instability. In the first approach, we approximate the beams in \cite{PozdeyevThesis} with elongated ellipses with uniform density. Elliptic beams of this type are rotating equilibria of Eq.\eqref{eq:nondimensional}, with angular velocity $\Omega = ab/(a+b)^2$ where $a$ is the semi-major axis of the ellipse, and $b$ its semi-minor axis \cite{Lamb}. Love has shown that when the aspect ratio of the ellipse exceeds the critical aspect ratio $(a/b)_{crit}=1/3$, wave-like disturbances on the vortex rim, known as Kelvin modes \cite{Balmforth}, become unstable. The elliptic approximations of the beams in \cite{PozdeyevThesis} exceed the aspect ratio threshold for instability. Mitchell and Rossi \cite{Mitchell} give in Equation 17 a formula to find the azimuthal mode number $m$ for the fastest growing mode at a given aspect ratio. For a 30 to 1 ellipse, the fastest growing modes are $m=13$ and $m=14$, which lead to the formation of 7 clusters \cite{Mitchell}. This is to be compared with the 9 clusters obtained in the PIC simulation of the equivalent beam in \cite{PozdeyevThesis}. This slight disagreement is expected, given that in \cite{PozdeyevThesis} the beam density is not uniform, and the beam is not elliptic in the median plane but instead rectangular.
The second way to characterize the instability is to consider the actual radial density distribution of the beam, but to approximate the beam as infinitely long. In Figure \ref{fig:shearlayer} we show the $\mathbf{E}\times\mathbf{B}$ field for the finite length beam in \cite{PozdeyevThesis}. The radial distribution of the beam density is Gaussian: $n(x)=\exp(-\beta x^2/h^2)$, where $h$ is the width of the beam, and $\beta=-4\ln(10^{-3})$. $n$ is independent of $y$ in the beam, but falls off abruptly to 0 at $y=\pm 15 h$. The reversal of the $\mathbf{E}\times\mathbf{B}$ shear flow inside the beam is visible in the figure. 
If we consider an infinitely long beam, we recover the classical situation of a two-dimensional parallel shear flow \cite{Drazin}. Solving Rayleigh's stability equation \cite{Drazin} numerically, we verify that the $\mathbf{E}\times\mathbf{B}$ shear flow due to our density distribution is unstable, and find that the fastest growing mode is such that $kh\approx 1.68$, where $k$ is the wave number of the instability. For $h=1$ cm as in the first numerical experiment in \cite{PozdeyevThesis}, the wavelength of the most unstable mode is $\lambda\approx 3.74$ cm according to our model, so that approximately 8 wavelengths fit in a 30cm long beam. Because the beam is in fact finite in the longitudinal direction, spiraling occurs for roughly half a wavelength at the top and bottom of the beam. This means that 9 clusters are initially formed: two due to spiraling and seven due to the shear layer instability. Our simulation shown in Figure \ref{fig:instability} supports these estimates, and recovers the features found in PIC simulations: 1) the initial spiraling of the head and tail of the beam due to the stronger $\mathbf{E}\times\mathbf{B}$ flow at the beam ends; 2) the development of the instability in the core of the beam, leading to the formation of clusters; 3) the roll up of the beam head and tail at the same time, which eventually merge with nearby clusters. The merging of vortex pairs is thought to be closely linked with axisymmetrization, and is a well documented process in fluid dynamics \cite{Winant,Meunier,Leweke}. By computing the beam evolution for longer times, we see that the merging of pairs of clusters continues, eventually leading to one large round beam at the end of the simulation. For the early phase of clustering, our model predicts that for a given density profile, the number of clusters formed is inversely proportional to the radial size of the beam, since the wave number is scaled by $h$. The model also predicts that the number of clusters is proportional to beam length, since it is determined by the number of wavelengths of the fastest growing mode that fit in the beam. Both predictions agree with published results \cite{Pozdeyev_PRSTAB,PozdeyevThesis}.

In summary, beam spiraling and beam break up are both consequences of the advection of the beam in the $\mathbf{E}\times\mathbf{B}$ velocity field, where $\mathbf{E}$ is the self electric field. This $\mathbf{E}\times\mathbf{B}$ dynamics is described by the two-dimensional Euler equations for an incompressible fluid. Spiraling is the beam equivalent of the axisymmetrization principle for isolated vortex distributions, and beam break up is the result of the shear layer instability. Our results demonstrate that 2-D fluid dynamics can guide the design and operation of high intensity cyclotrons. Conversely, our work suggests that cyclotrons could be used for the experimental study of isolated 2-D vortices with a decay boundary condition on the stream function at infinity, just like Penning-Malmberg traps are used for the study of vortices with no normal flow boundary condition at a wall \cite{Driscoll,Fine}. Cyclotrons offer the unique possibility to study the dynamics of inviscid shear layers experimentally by injecting very elongated beams. Likewise, despite likely difficulties with the control of the beam profile at injection, cyclotrons may be well suited for the first experimental verification of stationary V-states \cite{Deem,Burbea,Hmidi}, which would be uniform beams in rotating steady state with $2\pi/m$ rotational symmetry and $m>2$.

\section*{Acknowledgements}
The author would like to thank Thomas Planche (TRIUMF) for early discussions on beam break up in cyclotrons, and Oliver B\"uhler (NYU CIMS) for a very useful review of the relevant fluid dynamics literature.

\end{document}